\documentclass[twocolumn,showkeys,preprintnumbers,amsmath,amssymb]{revtex4}

\usepackage{graphicx}
\usepackage{dcolumn}
\usepackage{bm}
\usepackage{epsfig}

\begin{document}
\preprint{IST/CFP 6.2006}

\title[]{Electron trapping by electric field reversal in glow
discharges}

\author{Mario J. Pinheiro}
\affiliation{Department of Physics and Center for Plasma Physics,
Instituto Superior T\'{e}cnico, Av. Rovisco Pais,
 1049-001 Lisboa, Portugal}
\email{mpinheiro@ist.utl.pt}



\bibliographystyle{apsrev}

\begin{abstract}
The phenomena of electric field reversal in glow discharges is
discussed. Several models are described and the link between field
reversal and the discharge structure is analyzed.
\end{abstract}

\maketitle \texttt{}

\section{Introduction}

The phenomena of field reversal of the axial electric field in the
negative glow of a dc discharge has been discussed since a long
time in literature~\cite{Raizer1,Engel1}. In a founding paper
Druvestyn and Penning~\cite{Druyvesteyn} hypothesized its
existence. Lawler {\it et al.} ~\cite{Lawler1} assumed its
existence when evaluating the current balance at the surface of a
cold cathode discharge and estimated the position of the field
reversal by linearly extrapolating the optical data of atomic
transitions. The determination of the position where the field
reversal occurs is of great importance since the fraction of ions
returning to the cathode depends on its existence and location,
and it is necessary for correctly determine the conditions for a
self-sustained discharge.

Kolobov and Tsendin ~\cite{Kolobov:92} have shown that the first
field reversal is located near the end of the negative glow (NG),
near the position (although located slightly to the cathode side)
where ions density attains its greatest
magnitude~\cite{Gottscho:89}. If the discharge length has enough
extension and the pressure decrease to lower values it appears a
second field reversal on the boundary between the Faraday dark
space and the positive column (PC) ~\cite{Kolobov:92,Gottscho:89}.

Moreover, Kolobov and Tsendin explained how ions produced to the
left of the first reversal location move to the cathode by
ambipolar diffusion - helping to maintain the glow by secondary
electron emission - and ions generated to the right of this
location drift to the anode.

More recent work ~\cite{Maric:02,Maric:03} presenting a comparison
of experimental data and the predictions of a hybrid fluid-Monte
Carlo model also supports the view that the point where the field
is extrapolated to zero is practically coincident with the maximum
of the emission (even when $j/p^2$ scaling is no longer valid).
Those characteristics were experimentally observed by laser
optogalvanic spectroscopy~\cite{Gottscho:89}. For a detailed
review see also~\cite{Godyak}.

Boeuf and Pitchford~\cite{Boeuf} with a simple fluid model gave an
analytical expression of the field reversal location showing its
dependency solely on the cathode sheath length, the gap length,
and the ionization relaxation length. They obtained as well the
fraction of ions arriving at the cathode and the magnitude of the
plasma maximum density.

Technological application of gas discharges, particularly to
plasma display panels, needs a better knowledge of the processes
involved.

\section{Structure of a Glow Discharge}

The phenomenology of DC glow discharges are well described, for
example, by Roth~\cite{Roth:95}, where the existence of the field
reversal in the negative glow (NG) is already clearly shown. To be
complete, we will recall now the most significant features with
interest to our problem. In the regime of a {\it normal glow
discharge} the voltage and the current density are both
practically independent of the total current flowing in the
discharge tube.


In a crude picture, energy is continuously transferred from a high
voltage DC power source to the electrons originating from the
cathode that accelerate absorbing energy from the field $eE$,
ionizing, exciting and undergoing elastic collisions with heavy
particles and other electrons. Electrons disappear basically by
recombination and diffusion to the walls. Secondary electrons from
the cathode provides an additional source of electrons and they
have a prominent role in the discharge maintenance. The electric
field decreases in the sheath, vary slowly in the plasma and
increases again in the anode sheath, although not attaining such
high value as in the cathode. All plasmas are separated from the
walls (both conducting or non-conducting) by a sheath. The higher
energies are attained by electrons which bombard far more
frequently the walls than ions. The slower ions remaining behind
build-up a positive plasma potential relative to the wall with a
magnitude of the order of the electron kinetic temperature. In
consequence ions are easily drawn off through the sheath by an
external circuit  as a current $I$ while electrons are repelled by
the walls (and electrodes), only escaping the most energetic ones.
As we proceed from the cathode to the anode we recognize several
regions, first observed by Michael Faraday in the 1830's, with
interest to the present study (see also
Refs.~\cite{Roth:95,Laporte:39}):
\begin{itemize}
    \item Aston dark space - The electric field is very high
    and primary and secondary electrons outnumber ions. However,
    electrons have not still attained enough energy to excite
    neutrals and they have low density making this region to appear dark;
    \item cathode glow - electrons have already enough energy to
    excite neutrals and ions density increases;
    \item cathode (Crookes, Hittorf) dark space - Electrons are either too fast or too slow
    to excite the gas. The probability for fast electrons to excite are not negligible, however, and that's why
    this regions is slightly brighter than the Aston dark space.
    In this region electrons have enough energy to ionize the gas
    and the ions production rate attains high magnitude, the dominant
    charge being positive. It
    is in this region that the primary electrons loose a great fraction of its energy in
    ionizing collisions and the generated secondary electrons appear with lower energy. Slow
    electrons are produced in this region. A moderate electric
    field dominates;
    \item cathode region (CR) - This is a transition region between the cathode
    dark space and the negative glow where most of the voltage drop
    (cathode fall) of $\phi_c$ at $x=d_c$. The axial length of the
    cathode region $d_c$ adjust its value in order to maintain a
    minimum value of the product $d_c p$, the Paschen minimum;
    \item negative glow (NG) - The electric field is here very
    low, but its luminosity results from the electrons that have
    been accelerated in the CR and ionize and excite neutrals in
    the NG (that's why the luminosity is more intense on the CR
    side). As electrons lose their energy, recombinations processes are more
    probable, mainly in poliatomic electronegative gases to which
    electrons attach more easily. Electrons carry almost all the
    entire electric current in this region. Ionization in the glow
    can contribute significantly to the total ion current in the
    glow due to backscattered ions from the NG entering the cathode
    fall~\cite{Dalvie:92}.
    \item Faraday dark space (FDS) - This region is located immediately to
    the right of the NG. Electrons energy is low and their
    density decreases due to recombination and radial diffusion.
    However, in gas discharge tubes with a radius much smaller
    than the discharge length, the electric field start to grow
    again and electrons gain energy giving birth to the positive
    column.

\end{itemize}

\section{Experimental observation of field reversal}

The phenomenon of field reversal was experimentally observed by
laser optogalvanic spectroscopy (LOG) firstly by Gottscho {\it et
al.}~\cite{Gottscho:89}. The LOG technique used was reported by
Walkup {\it et al.}~\cite{Walkup:83} and is based on a change of
ion mobility on excitation by a dye laser. This means that an
excited ion state with a larger (smaller) mobility will induce a
transient current increase (decrease) on excitation. The sign of
the LOG signal depends only on the sign of electric field
reversal. Gottscho {\it et al.}~\cite{Gottscho:89} at the same
time recorded the $N_2^+ (X^2 \Sigma_g^+)$ ions excited from its
lowest and first excited vibrational levels (exciting the $B^2
\Sigma_u^+ - X^2 \Sigma_:g^+$ (0,0) and (0,1) bands, respectively)
using laser-induced fluorescence (LIF) along the plane parallel
electrodes of their reactor. They noted that at low pressure the
LOG signal changes sign near the position of the $N_2^+$ maximum
while being located slightly to the cathode side. At higher
pressure the field reversal occurs toward the anode side of the
$N_2^+$ density maximum.

The LOG and LIF spatial profiles resulting from the excitation of
the $B^2 \Sigma_u^+ - X^2 \Sigma_:g^+$ (0,0) and (0,1) bands
decrease first strongly to negative values due to the falloff in
ion density, as it is registered by LIF. This is the
pre-sheath~\cite{Roth:95}, a quasi-neutral region between the
plasma and the sheath characterized by small electric fields
gradients (usually smaller than 1 V$/$cm). At a given position
ions practically remain undetected by LIF, the LOG signal
diminishes rapidly and this is a signature of the beginning of the
cathode fall region.


\section{Physical mechanism of the electric field reversals in glow
discharges}

Electric field reversals operate as a self-regulator mechanism of
glow discharges. If electrons were in hydrodynamic equilibrium
with the local electric field and were created in the cathode
sheath followed by a fast diffusion to the walls and acceleration
to the anode, then ions loss would occur through migration to the
cathode and it will be no need for field reversal. This situation
would be similar if the ionization main channels were ion and fast
neutral impact ionization instead of electron impact ionization.
However, electrons are not in equilibrium with the local electric
field. This situation results from the complex structure embodied
in the sheath, pre-sheath and plasma regions (see
Ref.~\cite{Roth:95} for details). The electric field decreases
largely from the cathode surface within an adjacent region where
the charged-particle density is low (e.g., ~\cite{Hartog:88}). In
this region electrons gain energy being strongly accelerated
toward the bulk plasma attaining typically kinetic energies in the
range $10 \div 10^3$ eV. Thus, they are responsible of the
excitation and ionization processes in the quasi-neutral region we
call plasma. Clearly their energy is determined by the
electron-energy spectrum and not by the local value of the
electric field~\cite{Shi}.

Under typical gas discharge conditions it is possible to consider
two groups of electrons: the group of slow and fast electrons. The
majority of them belongs to the slow group of electrons; they are
created in the quasi-neutral (bulk) plasma region through
ionization and their kinetic energy does not exceed significantly
the energy of the first excitation level
$\epsilon_1$~\cite{Tsendin95}. The low electric field prevalent in
the quasi-neutral region with a vast majority of slow electrons
(with energies in the range $10^{-2} \div 10$ eV) is favorable to
increase electrons loss while at the same time reducing ions loss.
But as the total current density
\begin{equation}\label{eq1}
\mathbf{j}=\mathbf{j}_e + \mathbf{j}_i = -en_e v_e + e n_i v_i
\end{equation}
must be constant along the discharge a mechanism of field reversal
appears in order to enhance ions loss and reduce slow electrons
loss. At the position where field reversal takes place the plasma
potential presents a minimum value~\cite{Gottscho:89}. This
situation creates a kind of two contiguous {\it reservoirs}: to
the left of the field reversal, ions generated by fast electrons
flow to the cathode, where hitting the electrode surface induce
secondary electrons emission at the cathode (and thus helping
maintaining the glow); to the right of the field reversal, ions
are accelerated toward the anode. As electrons diffuse more
rapidly to the walls than ions, the walls along the positive
column and near the anode sheath charge negatively, while along
the cathode fall they charge positively (see, for example, Ref.
~\cite{Laporte:39}). This charge build-up deform the
equipotentials surfaces that, instead to remain parallel to the
electrodes, show the concave side facing the anode.

\section{Kolobov and Tsendin nonlocal kinetic model}

Due to the complex self-consistent problem that is at stake,
Kolobov and Tsendin endeavored to develop an analytical approach
with a formidable complexity. Since the papers of Emeleus and
co-workers ~\cite{Emeleus:34,Emeleus:39} electrons have been
divided in primary, secondary (or intermediate) and ultimate (or
trapped) and so did Kolobov and Tsendin. The highly energetic
electron created in the cathode sheath of a DC discharge have
kinetic energies well above the first excitation potential
$\varepsilon^{*}$. Ionization and excitation processes are created
by them and they carry the electron current in the sheath and
nearby plasma region. The plasma region is mainly populated by
trapped electrons that do not contribute to the current. In the NG
and FDS the current is mainly carried by (untrapped) intermediate
electrons with energies below $\varepsilon^{*}$. The discharge gap
is divided in two regions: the sheath and the (quasi-neutral)
plasma region.

The fast electrons are described by a continuous-energy-loss model
neglecting scattering
\begin{equation}\label{eq2}
\frac{dw}{d \xi} = -N \mathcal{L}(w) - e E (x),
\end{equation}
where $w$ is the electron kinetic energy, $N$ is the neutral
particle density, $\mathcal{L}(w)$ is the energy-loss function,
$\xi$ is the fast-electron path along its trajectory and $x$ is
the spatial coordinate.

Neglecting scattering, then $\xi=x$ and the kinetic equation for
fast electrons $F(v,x)$ is written under the form
\begin{equation}\label{eq3}
v \frac{\partial F}{\partial x} - \frac{eE(x)}{m} \frac{\partial
F}{\partial v} - \frac{\partial }{\partial v} \frac{N
\mathcal{L}(w)}{m} F = J(x,v).
\end{equation}
Here, $J(x,v)$ is a source term and it was assumed $\mathcal{L}(w)
= \mathcal{L}_0 = const$. Its integration is performed at constant
energy
\begin{equation}\label{eq4}
\varepsilon = w + N \mathcal{L}_0 x - e \phi(x) = w - e
\tilde{\phi}(x),
\end{equation}
where $\phi(x)$ is the electrostatic potential obtained for
linear-electric field profile in the sheath. The fast electrons
start riding on an effective potential, $e \tilde{\phi}(x)$, that
attains a minimum in a point where the total force is zero, $eE
(\tilde{d})=N\mathcal{L}_0$ (see Fig.2). For example, for He,
$\mathcal{L}_0=1.5 \times 10^{-15}$ eV cm$^2$. Solving
Eq.~\ref{eq3} it is obtained the fast-electron current
\begin{displaymath}\label{eq5}
j_f(x)=e \int_0^{\infty} v dv F(v,x)=
\end{displaymath}
\begin{equation}\label{eq6}
j_f(x) = \left \{ \begin{array}{cc}
  e \Gamma_e \exp(\alpha x), & 0<x<\tilde{d} \\
  j_f(x_0(\varepsilon=e\mathcal{\phi}(x))), & \tilde{d} < x < x_1(e \tilde{\phi}_c),
\end{array}
\right.
\end{equation}
where $\alpha=N \mathcal{L}_0/\varepsilon_0$, with $\varepsilon_0$
denoting the constant energy loss per ion-electron pair and
$\Gamma_e$ is the electron flux from the cathode. Notice that for
$x < \tilde{d}$ the total electron current $j_e$ is equal to the
fast-electron current $j_f$ and that, in particular, we have
\begin{equation}\label{eq7}
\frac{d j_e}{d x} = \alpha j_f = I(x).
\end{equation}
Here, $I(x)$ is the ionization density. Eqs.~\ref{eq6} and
~\ref{eq7} constitute a generalization of the Townsend approach
with a constant ionization coefficient $\alpha=const.$

\begin{figure}
  \includegraphics[width=3.75 in, height=4 in]{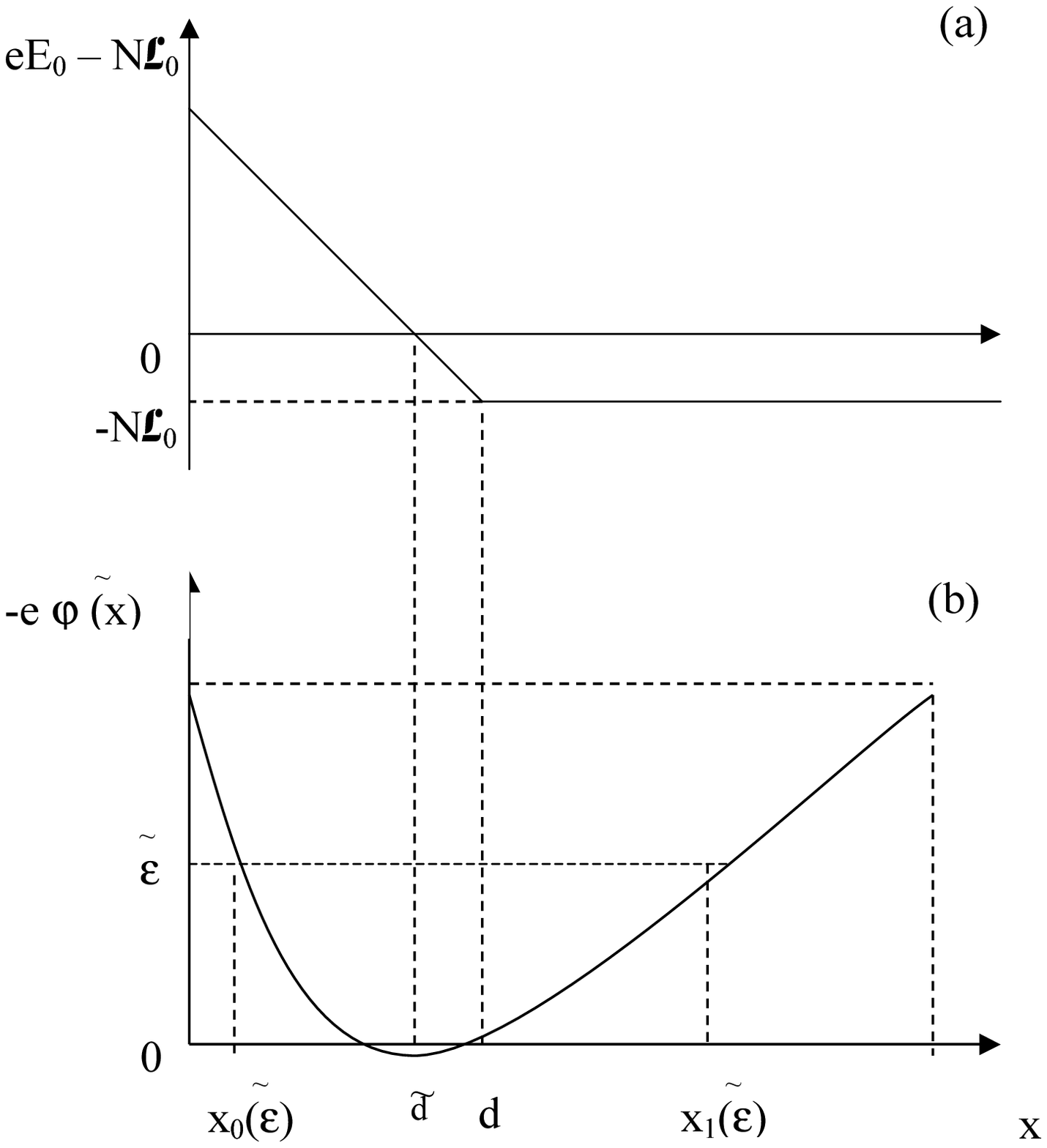}\\
  \caption{a) Total force acting on an electron under constant energy-loss; b) effective potential energy of
  electrons, $e \tilde{\phi}(x)$)}\label{fig2}
\end{figure}

At $x > x_1(e \tilde{\phi}_c)$ there is no ionization source since
the fast electrons do not reach this region; this region limits
the boundary between the FDS and NG (see Sec. 2 for its
characterization) and the length of the NG is given qualitatively
by $R=e \phi_c/\alpha \varepsilon_0$.

For slow electrons with $w < \varepsilon^{*}$ the
continuous-energy-loss is no longer valid and it is essential to
take into account the essential discrete character of the energy
loss, where elastic and electron-electron collisions are the only
energy-relaxation mechanisms. If $L$ is bigger than the electron
mean free path $\lambda$, then the electron energy distribution
function (EDF) is isotropic and the two-term expansion in
spherical harmonics holds. Hence, taking into account the
following contributions: i) degradation of fast electrons injected
into the plasma from the cathode sheath; ii) slow electrons
generation by fast electrons; iii) slow electrons superelastic
collisions with molecular low vibrational levels or others, with
kinetic energy slightly exceeding $\varepsilon^{*}$, the kinetic
equation for the isotropic part of the EDF gives for the trapped
electrons a Maxwell-Boltzmann distribution of the type:
\begin{equation}\label{eq8}
f_0^{(t)}=\left( \frac{m}{2 \pi T_e} \right)^{2/3} n_m [\exp
(-\varepsilon/T_e) -\exp(-e \phi_a/T_e)],
\end{equation}
where $\phi_a$ denotes the anode potential and $n_m$ is the
maximum trapped electron density at the point of field reversal
$E(x)=0$. The electron temperature is obtained by solving the
integral energy balance for trapped electrons, but in a first
approach we have $e \phi_a \sim T_e$. For example, by laser
diagnostics experiments~\cite{Hartog:88} the slow electron
temperature and density are $T_e=0.2$ eV and $n_m=1.2 \times
10^{12}$ cm$^3$in He at $p=3.5$ Torr, with an electrode gap
$L=0.62$ cm and $j=0.846$ mA$/$ cm$^2$. The trapped electrons give
the main contribution to the plasma density
\begin{equation}\label{eq9}
n_s(x)=\frac{4 \pi}{m} \int_{-e\phi(x)}^{\varepsilon^{*}}
v(\varepsilon,x) f_0(\varepsilon,x) d\varepsilon,
\end{equation}
while giving zero contribution to the slow-electrons current
density:
\begin{equation}\label{eq10}
j_s (x) = \frac{4 \pi e}{3m} \int_{-e\phi(x)}^{\varepsilon^{*}}
v(\varepsilon,x) D(\varepsilon,x) \frac{\partial f_0}{\partial x}
d\varepsilon.
\end{equation}
The $j_s$ is the the total electron current in the FDS represented
by the transport of the intermediate untrapped electrons. In the
low-pressure limit the trapped electrons are expected to be
isothermal and to be governed by the ambipolar diffusion equation:
\begin{equation}\label{eq11}
\frac{d}{dx} \left( D_a \frac{dn}{dx} \right) +
I(x)-\frac{n}{\tau}=0.
\end{equation}
Here, $\tau = \Lambda^2/D_a$. Eq.~\ref{eq11} can be solved using
the Green's function approach with appropriate boundary
conditions. It gives the minimum $x_m$:
\begin{equation}\label{eq12a}
\int_d^L xI(x) dx = (L + d) \int_{x_m}^L I(x) dx.
\end{equation}
The position of $x_m$ determines the ratio of ions returning to
the cathode. If the FDS length $(L-R)$ exceeds $(R-d)$ then $x_m
\to R$ and the majority of ions generated in the plasma return to
the cathode; if, on the contrary, $R-d > L-R$ this means that fast
electrons penetrate deep in the plasma, the majority of them
attaining the anode. In their calculation the position of maximum
ions density coincides with the field reversal location.
Unfortunately, Kolobov and Tsendin do not give explicit expression
for $x_m$, introducing an error in Eq.~\ref{eq12a} (see remark in
Ref.~\cite{Kudryavtsev:05}).

To resume the previous work by Kolobov and
Tsendin~\cite{Kolobov:92}, the study of nonlocal phenomena in
electron kinetics of collisional gas discharge plasma have shown
that in the presence of field reversals the bulk electrons in the
cathode plasma are clearly separated in two groups of slow
electrons: trapped and free electrons. Trapped electrons give no
contribution to the current but represent the majority of the
electron population. The first field reversal it was shown
qualitatively to be located near the end of the negative glow (NG)
where the plasma density attains the greatest magnitude. If the
discharge length is long enough, it appears a second field
reversal on the boundary between the Faraday dark space and the
positive column. Also, it was shown that ions produced to the left
of this first reversal location move to the cathode by ambipolar
diffusion and ions generated to the right of this location drift
to the anode. For a review see also~\cite{Godyak}.

\section{Boeuf and Pitchford fluid model}

Boeuf {\it et al}~\cite{Boeuf} with a simple fluid model gave an
analytical expression of the field reversal location which showed
to depend solely on the cathode sheath length, the gap length, and
the ionization relaxation length. They obtained as well a simple
analytical expression giving the fraction of ions returning to the
cathode and the magnitude of the plasma maximum density.

\begin{figure}
  \includegraphics[width=3.5 in, height=3 in]{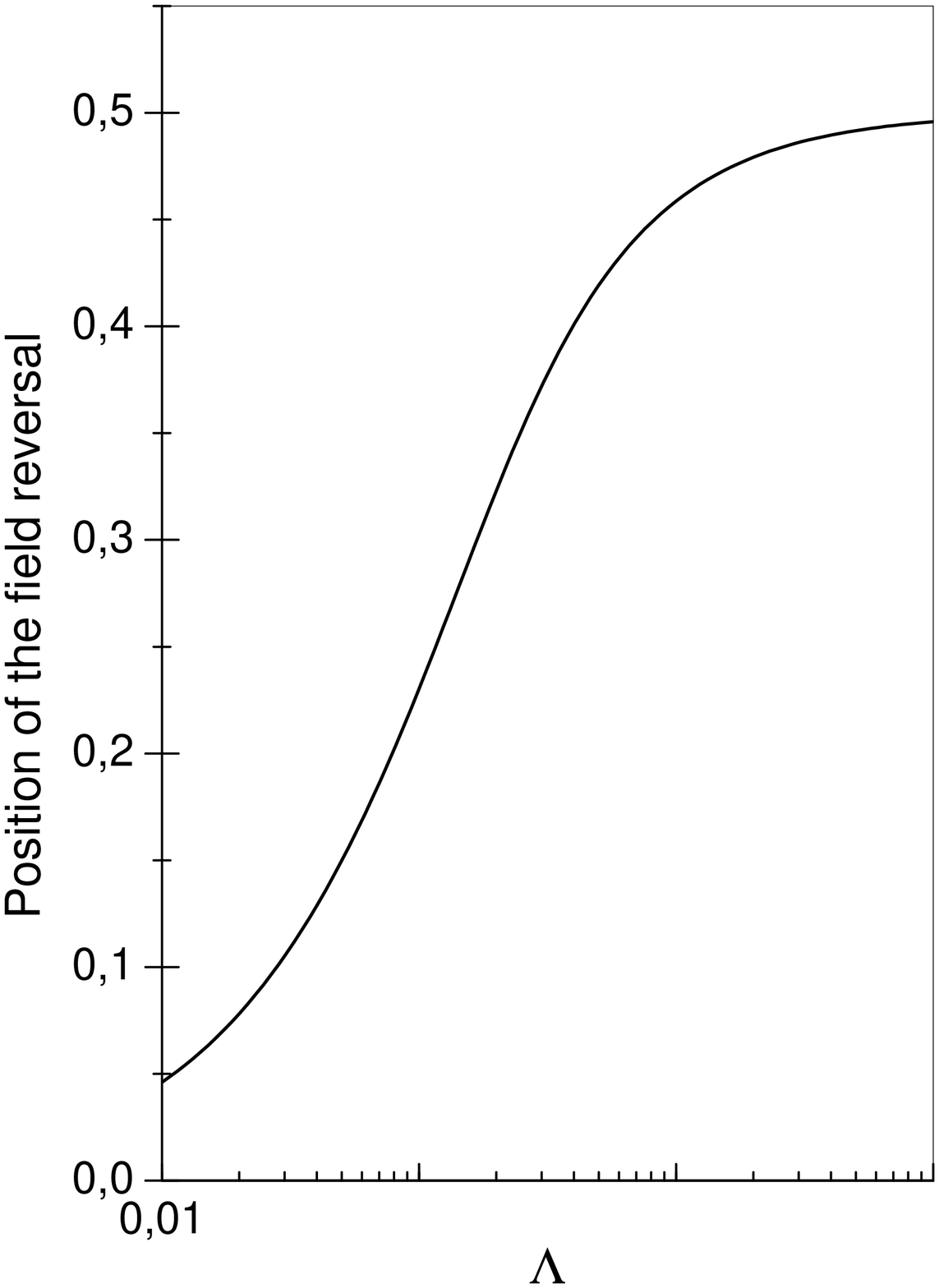}\\
  \caption{Position of the field reversal, $\frac{x_0 - d_c}{L - d_c}$, as a function of the
  adimensional parameter $\Lambda$.}\label{fig3}
\end{figure}

Their model assumes a spatial profile of the ionization rate
$S_T(x)=N_e(x) \nu_i (x)$ in a plasma being formed in a parallel
plate geometry. Monte Carlo calculations show that the ionization
rate is linear in the sheath and decreases exponentially in the
NG~\cite{Peres:92} and so they assumed
\begin{equation}\label{eq13}
S_T (x) = \left\{ \begin{array}{ll}
  0                        & \mbox{,for $x < d_c$} \\
  s \exp(-\frac{x-d_c}{\lambda_{\varepsilon}}) & \mbox{,for $x \geq d_c$}.
\end{array}
\right.
\end{equation}
Here, $\lambda_{\varepsilon}$ is the energy relaxation length of
the high electrons energy entering the NG accelerated in the
sheath while $s$ is a function of the voltage $\phi_c$, gas
mixture and is proportional to the flux of secondary electrons
leaving the cathode $\Gamma_{e0}$ (here, $e$ (electrons) and $i$
(ions) denotes the specie). They assume that slow electrons can be
described by a continuity equation and a momentum transport
equation in the drift-diffusion approximation (and thus
considering drift energy negligible with respect to thermal
energy):
\begin{equation}\label{eq14}
\frac{d \Gamma_{e0}}{dx} = S_T(x),
\end{equation}
with
\begin{equation}\label{eq15}
\Gamma_{e0} = \mp n_{e,i} \mu_{e,i} E - D_{e,i} \frac{d
n_{e,i}}{dx},
\end{equation}
where $\mu_{e,i}$ and $D_{e,i}$ denote the electron and ion
mobility and diffusion coefficients respectively. Quasi-neutrality
($n=n_e \sim n_i$) is assumed in the NG region. The total current
density is given by the expression
\begin{equation}\label{eq15a}
J_T =e (-\phi_e + \phi_i).
\end{equation}
From this equation, together with Eq.~\ref{eq15}, it is obtained the
electric field
\begin{equation}\label{eq15b}
E(x) = \frac{J_T}{e n_e (\mu_e + \mu_i)} - \frac{D_e - D_p}{\mu_e
+ \mu_i}\frac{1}{n}\frac{dn}{dx},
\end{equation}
where the electron and ion fluxes are now given by
\begin{equation}\label{eq15c}
\phi_e (x) = - \frac{J_T \mu_e}{e(\mu_e + \mu_i)} - D_a
\frac{dn}{dx},
\end{equation}
and
\begin{equation}\label{eq15d}
\phi_i (x) = - \frac{J_T \mu_i}{e(\mu_e + \mu_i)} - D_a
\frac{dn}{dx}.
\end{equation}
Here, $D_a$ is the ambipolar diffusion coefficient given by
\begin{equation}\label{eq15e}
D_a = \frac{\mu_e D_i + \mu_i D_e}{\mu_e + \mu_i}.
\end{equation}
Integrating Eq.~\ref{eq14} and taking into account the ionization
source terms in Eq.~\ref{eq13} it is obtained
\begin{equation}\label{eq16}
\Gamma_e (x) = \left\{ \begin{array}{ll}
  \Gamma_{e0}                        & \mbox{,for $x < d_c$} \\
  \Gamma_{e0} + s \lambda_{\varepsilon}[1-a(x)] & \mbox{,for $x \geq
  d_c$},
\end{array}
\right.
\end{equation}
using the simplifying notation
$a(x)=\exp[-(x-d_c)/\lambda_{\varepsilon}]$. At some position
$x_m$ it can possibly occur a field reversal, located at a
boundary that split the ion flux at the right side toward the
anode and keeping electrons confined in the plasma region while at
the left side directing the ion flux toward the cathode. Hence, a
condition can be stated of a total current at some yet unspecified
position $x_0$ (in this Sec. we use this notation for the field
reversal location) being equal to the electron current:
\begin{equation}\label{eq17}
J_T = -e \Gamma_e(x_0) = -e \{\Gamma_{e0} + s
\lambda_{\varepsilon}[1-a(x_0)] \}.
\end{equation}
From Eq.~\ref{eq15c} and Eq.~\ref{eq16}, using Eq.~\ref{eq17}, it
is obtained the following first order differential equation for $x
\geq d_c$:
\begin{equation}\label{eq18}
1 + \frac{s \lambda }{\phi_{e0}}[1-a(x)]=\frac{\mu_e}{(\mu_e +
\mu_i)} \left( 1 + \frac{s\lambda}{\phi_{e0}}[1-a(x_0)] -
\frac{D_a}{\phi_{e0}} \frac{dn}{dx} \right).
\end{equation}
Integrating Eq.~\ref{eq18} between $d_c$ and $L$ with appropriate
boundary conditions, $n(d_c)=n(d)=0$, under the approximation
$\mu_i \ll \mu_e$, it is obtained
\begin{equation}\label{eq19}
a(x_0)=\frac{\lambda}{d - d_c} [1-a(d)].
\end{equation}
Introducing the dimensionless parameter $\Lambda = \lambda /(d -
d_c)$, representing the ratio of the relaxation length to the
distance between the anode and the CR boundary, Eq.~\ref{eq19} can
be written under the form
\begin{equation}\label{eq20}
\frac{x_0 - d_c}{L - d_c} = - \Lambda \ln [ \Lambda (1- \exp
(-1/\Lambda)].
\end{equation}
The above equation shows that, by one side, the position of the
field reversal $x_0$ depends on the cathode sheath length $d_c$,
the gap length $L$ and on the electron energy relaxation length
$\lambda_{\varepsilon}$. By other side, $d_c$ and
$\lambda_{\varepsilon}$ depend on the cathode sheath voltage and
gas mixture. As can be seen in Fig.~\ref{fig3}, when
$\lambda_{\varepsilon} \ll L$ (low values of $\Lambda$) the
position of the field reversal is near the plasma-sheath boundary;
when $\lambda_{\varepsilon} \gg L$ (large values of $\Lambda$) the
position of the field reversal is displaced to the mid-distance
between the gap length and the plasma-sheath boundary. This last
case occurs for an obstructed discharge, i.e, when the gap length
is less than the Paschen minimum $d_c$ at the Paschen minimum.

\begin{figure}
  \includegraphics[width=3.5 in, height=3 in]{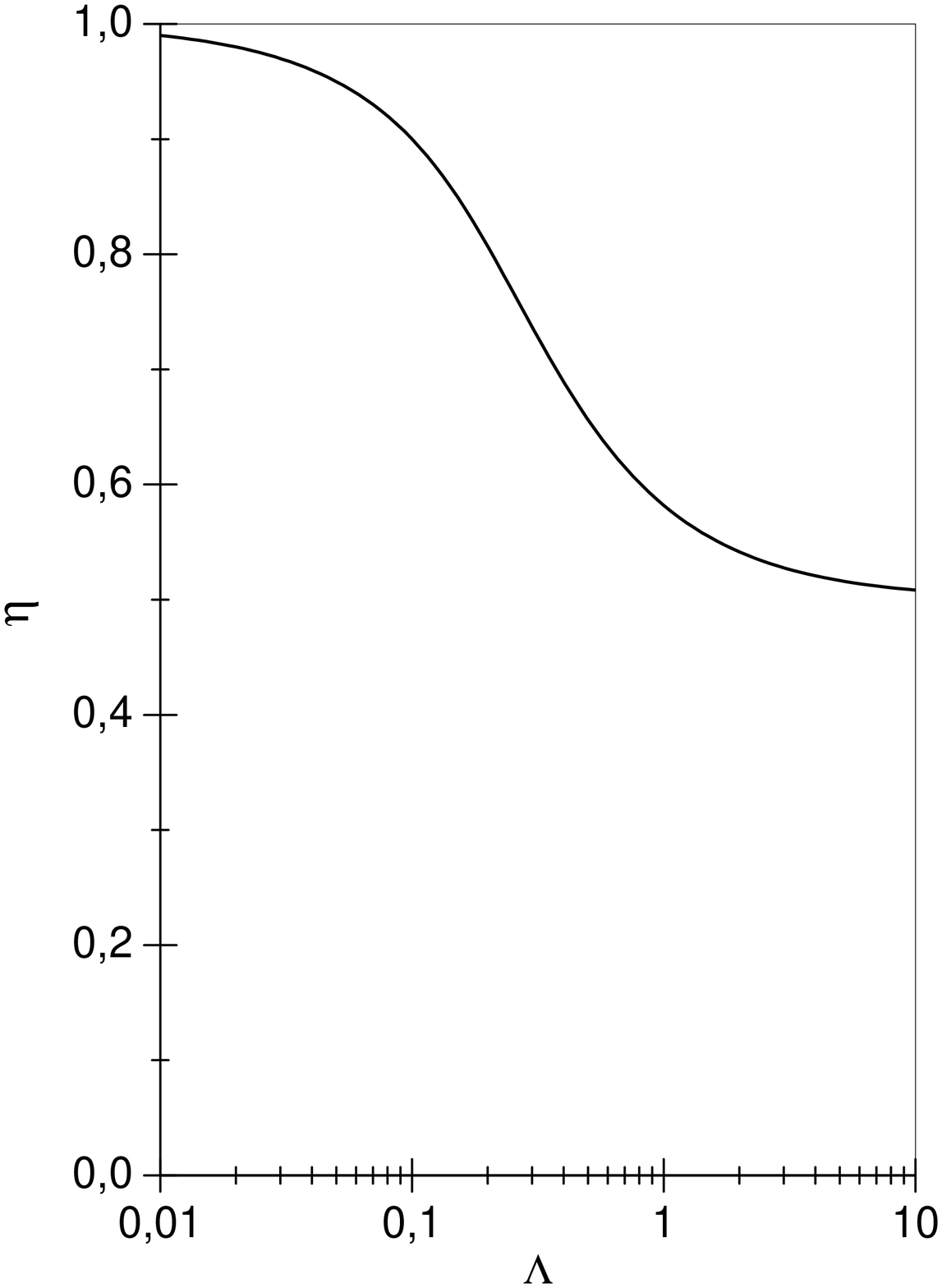}\\
  \caption{Fraction of ions returning to the cathode as a function of the
  adimensional parameter $\Lambda$.}\label{fig4}
\end{figure}

The fractions of ions to the cathode can be determined using the
relationship
\begin{equation}\label{eq21a}
\eta = \frac{\int_0^{x_0} S_T(x) dx}{\int_0^L S_T(x)dx} =
\frac{\phi_e(x_0) - \phi_{e0}}{\phi_e(L) - \phi_{e0}}.
\end{equation}
We can give an explicit form to this ratio using Eq.~\ref{eq16}
and Eq.~\ref{eq20}:
\begin{equation}\label{eq22}
\eta = \frac{1}{1 - \exp(-\/\Lambda)} - \Lambda.
\end{equation}
This fraction depends only on $\Lambda$ and, as Fig.~\ref{fig4}
illustrate, its values are between $0.5$ and $1$.

Fig.~\ref{fig4} shows that when the electron energy relaxation
length $\lambda_{\varepsilon}$ is smaller than the gap length,
$\eta \to 1$, the major fraction of ions return to the cathode.
This is true if radial diffusion to the walls or recombination is
not accounted for. In the condition of an obstructed discharge
$\lambda_{\varepsilon} \gg L$, then $\eta \to 0.5$. Taking into
account radial diffusion and recombination will lower the fraction
of ions returning to the cathode.

Integrating Eq.~\ref{eq18} between $x=d_c$ and a given $x$,
considering appropriate boundary conditions, $n(d_c)=n(L)=0$,
assuming $\mu_i \ll \mu_e$ and using Eq.~\ref{eq19} it is obtained
readily
\begin{equation}\label{eq21}
n(x)=\frac{s \lambda_{\varepsilon}^2}{D_a} (1 - \exp(-1/\Lambda))
\left( \frac{1 - \exp(-\chi/\Lambda)}{1- \exp(-1/\Lambda)} -
\chi\right),
\end{equation}
with $\chi=\frac{x-d_c}{L-d_c}$. The plasma density attains a
maximum at the position where occurs the field reversal $x_0$.
This is inconsistent with Eq.~\ref{eq15b}, however, since here it
was neglected the ions mobility in regard of the electrons
mobility. In fact, the maximum of the plasma density is located
near the position of the field reversal, but its maximum should be
located to the cathode side of the field reversal. Maric {\it et
al.}~\cite{Maric:02,Maric:03} showed a perfect agreement of the
analytical formula given in Eq.~\ref{eq20} with an hybrid Monte
Carlo-fluid model. Kudryavtsev and Toinova~\cite{Kudryavtsev:05}
determined the position of the field reversal from the position of
maximum ion density in the plasma. Their approach is quite simple
giving simple analytical formulas in a short (without positive
column) DC glow discharge making use of the ionization rate
coefficients for fast electrons in the field.

\section{A Dielectric-like model of field reversal}

In this section we introduce a quite simple dielectric-like model
of a plasma-sheath system~\cite{Pinheiro:04}. This approach have
been addressed by other authors~\cite{Taillet69, Harmon76} to
explain how the electrical field inversion occurs at the interface
between the plasma sheath and the beginning of the negative glow.
The aim of this Letter is to obtain more information about the
fundamental properties related to field inversion phenomena in the
frame of a dielectric model. It is obtained a simple analytical
dependence of the axial location where field reversal occurs in
terms of macroscopic parameters. In addition, it is obtained the
magnitude of the minimum electric field inside the through, the
trapped well length, and the trapping time of the slow electrons
into the well. We emphasize in particular the description of the
dielectric behavior and do not contemplate plasma chemistry and
plasma-surface interactions.

The analytical results hereby obtained could be useful for hybrid
fluid-particle models (e.g., Fiala {\it et al.}~\cite{Boeuf94}),
since simple criteria can be applied to accurately remove
electrons from the simulations.

On the ground of the stress-energy tensor considerations it is
shown the inherent instability of the field inversion sheath. The
slow electrons distribution function is obtained assuming the
Fermi~\cite{Fermi} mechanism responsible for their acceleration
from the trapping well.

Lets consider a plasma formed between two parallel-plate
electrodes due to an applied dc electric field. We assume a planar
geometry, but extension to cylindrical geometry is
straightforward. The applied voltage is $V_a$ and we assume the
cathode fall length is $l$ and the negative glow + eventually the
positive column extends over the length $l_0$, such that the total
length is $L=l + l_0$. We have
\begin{equation}\label{Eq1}
-V_a = l E_s + l_0 E_p,
\end{equation}
where $E_s$ and $E_p$ are, resp., the electric fields in the
sheath and NG (possibly including the positive column).

At the end of the cathode sheath it must be verified the following
boundary condition by the displacement field $\mathbf{D}$
\begin{equation}\label{Eq2}
\mathbf{n}.(\mathbf{D}_p - \mathbf{D}_s) = \sigma.
\end{equation}
Here, $\sigma$ is the surface charge density accumulated at the
boundary surface and $\mathbf{n}$ is the normal to the surface. In
more explicit form,
\begin{equation}\label{Eq3}
\varepsilon_p E_p - \varepsilon_s E_s = \sigma.
\end{equation}
Here, $\varepsilon_s$ and $\varepsilon_p$ are, resp., the
electrical permittivity of the sheath and the positive column. We
have to solve the following algebraic system of equations
\begin{equation}\label{Eq5}
\begin{array}{cc}
  l_0 E_p + l E_s & = - V_a, \\
  \varepsilon_p E_p - \varepsilon_s E_s & = \sigma. \\
\end{array}
\end{equation}
They give the electric field strength in each region
\begin{equation}\label{Eq6}
\begin{array}{cc}
  E_s =  & -\frac{V_a}{L} \left(1-\alpha + \frac{l_o \sigma}{V_a \varepsilon_s}\right)\frac{1}{1-\frac{l\alpha}{L}}, \\
  E_p =  & -\frac{V_a}{L} \left(1-\frac{l \sigma}{V_a \varepsilon_s}  \right) \frac{1}{1-\frac{l\alpha}{L}}.\\
\end{array}
\end{equation}
Here, we define
$\alpha=1-\frac{\varepsilon_p}{\varepsilon_s}=\frac{\omega_p^2}{\nu_{en}^2}$.
Recall that in DC case,
$\varepsilon_p=1-\frac{\omega_p^2}{\nu_{en}^2}$, and
$\varepsilon_s=\varepsilon_0$, with $\omega_p$ denoting the plasma
frequency and $\nu_{en}$ the electron-neutral collision frequency.
In fact, our assumption $\varepsilon_s=\varepsilon_0$ is plainly
justified, since experiments have shown the occurrence of a
significant gas heating and a corresponding gas density reduction
in the cathode fall region, mainly due to symmetric charge
exchanges processes which lead to an efficient conversion of
electrical energy to heavy-particle kinetic energy and thus to
heating~\cite{Hartog:88}.

Two extreme cases can be considered: {\bf i}) $\omega_p
> \nu_{en}$, implying $\varepsilon_p < 0$, meaning that
$\tau_{coll} > \tau_{plasma}$, i.e, non-collisional regime
prevails; {\bf ii}) $\omega_p < \nu_{en}$, $\varepsilon_p > 0$,
and then $\tau_{coll}
> \tau_{plasma}$, i.e, collisional regime dominates.

From the above Eqs.~\ref{Eq6} we estimate the field inversion
should occurs for the condition $1-\frac{l_o \alpha}{L}=0$, which
give the position on the axis where field inversion occurs:
\begin{equation}\label{Eq8}
\frac{l_o}{L} = \frac{\nu_{en}^2}{\omega_p^2}.
\end{equation}

From Eq.~\ref{Eq8} we can resume a criteria for field reversal: it
only occurs in the non-collisional regime; by the contrary, in the
collisional regime and to the extent of validity of this simple
model, no field reversal will occur, since the slow electrons
scattering time inside the well is higher than the the well
lifetime, and collisions (in particular, coulombian collisions)
and trapping become competitive processes. A similar condition was
obtained in~\cite{Nishikawa69} when studying the effect of
electron trapping in ion-wave instability. Likewise, a
self-consistent analytic model~\cite{Kolobov:92} have shown that
at at sufficiently high pressure, field reversal is absent.

Due to the accumulation of slow electrons after a distance
$\xi_c=L-l_0$, real charges accumulated on a surface separating
the cathode fall region from the negative glow. Naturally, it
appears polarization charges on each side of this surface and a
double layer is created with a surface charge $-\sigma_1' < 0$ on
the cathode side and $\sigma_2'$ on the anode side. But, $\sigma'
= (\mathbf{P} \cdot \mathbf{n})$,
$\mathbf{P}=\mathbf{\mathbf{\varepsilon}_0} \chi_e \mathbf{E}$
with $\varepsilon = \varepsilon_0 (1 + \chi_e)$, $\chi_e$ denoting
the dimensionless quantity called electric susceptibility. As the
electric displacement is the same everywhere, we have
$\mathbf{D}_0 = \mathbf{D}_1 = \mathbf{D}_2$. Thus, the residual
(true) surface charge in between is given by
\begin{equation}\label{Eq9}
\sigma = - \sigma_1' + \sigma_2'.
\end{equation}
After a straightforward but lengthy algebraic operation we obtain
\begin{equation}\label{Eq10}
\sigma = \varepsilon_p V_a \frac{B}{A},
\end{equation}
where
\begin{equation}\label{Eq11}
A = L \left( -1+ \frac{\varepsilon_0 -
\varepsilon_s}{\varepsilon_p} \right) + l\left( -
\frac{\varepsilon_p}{\varepsilon_s} +
\frac{\varepsilon_s}{\varepsilon_p} \right),
\end{equation}
and
\begin{equation}\label{Eq12}
B = \frac{\varepsilon_0 (\varepsilon_s - \varepsilon_p)}{
\varepsilon_s \varepsilon_p }.
\end{equation}
We can verify that $\sigma$ must be equal to
\begin{equation}\label{Eqsig}
\sigma = \alpha \frac{V_a \varepsilon_0}{2 l_0}.
\end{equation}
Considering that $\sigma = \varepsilon_0 \chi_e E$, we determine
the minimum value of the electric field at the reversal point:
\begin{equation}\label{Eq13}
E_m = \frac{\omega_p^2}{\nu_{en}^2} \frac{V_a}{2 l_0 \chi_e}.
\end{equation}

Here, $\chi_e=\varepsilon_{rw}-1$, with $\varepsilon_{rw}$
designating the relative permittivity of the plasma trapped in the
well. From the above equation we can obtain a more practical
expression for the electrical field at its minimum strength
\begin{equation}\label{Eq14}
E_m = -
\frac{n_{ep}}{n_{ew}}\frac{\nu_{enw}^2}{\nu_{en}^2}\frac{V_a}{e
l_0} \approx - \frac{n_{ep}}{n_{ew}} \frac{T_{ew}}{T_{ep}}
\frac{V_a}{2 l_0}.
\end{equation}
The magnitude of the minimum electric field depends on the length
of the negative glow $l_0$. This also means that without NG there
is no place for field reversal, and also the bigger the length the
minor the electric field. The length of the negative glow can be
estimated by the free path length $l_0$ of the fastest electrons
possessing an energy equal to the cathode potential fall value
$eV_a$:
\begin{equation}\label{}
l_0 = \int_{0}^{eV_a} \frac{d w}{(N F(w))}.
\end{equation}
Here, $w$ is the electrons kinetic energy and $NF(w)$ is the
stopping power. For example, for He, it is estimated $p l_0=0.02
eV_a$ ~\cite{Kolobov:92} (in cm.Torr units, with $V_a$ in Volt).
We denote by $n_{ew}$ the density of trapped electrons and by
$T_{ew}$ their respective temperature. Altogether, $n_{ep}$ and
$T_{ep}$ are, resp., the electron density and electron temperature
in the negative glow region.

By other side, we can estimate the true surface charge density
accumulated on the interface of the two regions by the expression
\begin{equation}\label{Eq15}
\sigma = \frac{Q}{A} = - \frac{n_{ep} e A \Delta \xi}{A}.
\end{equation}
Here, $Q$ is the total charge over the cross sectional area where
the current flows and $\Delta \xi$ is the width of the potential
well.

\subsection{Instability and width of the potential well}

From Eqs.~\ref{Eqsig} and ~\ref{Eq15} it is easily obtained the
trapping well width
\begin{equation}\label{Eq16}
\Delta \xi = - \frac{e V_a}{2 m l_0 \nu_{enw}^2}.
\end{equation}
It is expected that the potential trough should have a
characteristic width of the order in between the electron Debye
length ($\lambda_{De}=\sqrt{\frac{\varepsilon_0 kT_e}{n_e e^2}}$)
and the mean scattering length. Using Eq.~\ref{Eq16}, in a He
plasma and assuming $V_a=1$ kV, $l_0=1$ m and $\nu_{en}=1.85
\times 10^{9}$ s$^{-1}$ (with $T_e=0.03$ eV) at 1 Torr ($n=3.22
\times 10^{16}$ cm$^{-3}$) we estimate $\Delta \xi \approx 2.6
\times 10^{-3}$ cm, while the Debye length is $\lambda_{De}=2.4
\times 10^{-3}$ cm. So, our Eq.~\ref{Eq16} gives a good order of
magnitude for the potential width, which is expected to be in fact
of the same order of magnitude than the Debye length.

Table I present the set of parameters used to obtain our
estimations. We give in Table II the estimate of the minimum
electric field attained inside the well. The first field reversal
at $\xi_c \approx l_{NG}$ corresponds to the maximum density
$n_{ew} \gg n_{ep}$~\cite{Boeuf,Tsendin2001}. So, the assumed
values for the ratio of electron temperatures and densities of the
trapped electrons and electrons on the NG are typical estimates.

\begin{widetext}
\begin{table}
  \centering
  \caption{Data used for $E/p=100$ V$/$cm$/$Torr. Cross sections and electron
  temperatures are
  taken from Siglo Data base, CPAT and Kinema Software, http://www.Siglo-Kinema.com }\label{Table1}
  \begin{tabular}{|c|c|c|}
    \hline
    Gas & $T_e$ (eV) & $\sigma$ ($10^{-16}$ cm$^{2}$)\\
    \hline
    Ar & 8 & 4.0 \\
    He & 35 & 2.0 \\
    O$_2$ & 6 & 4.5 \\
    N$_2$ & 4 & 9.0 \\
    H$_2$ & 8 & 6.0 \\
    \hline
  \end{tabular}
\end{table}
\end{widetext}

It can be shown that there is no finite configuration of fields
and plasma that can be in equilibrium without some external
stress~\cite{Longmire}. Consequently, this trough is dim to be
unstable and burst electrons periodically (or in a chaotic
process), releasing the trapped electrons to the main plasma. This
phenomena produces local perturbation in the ionization rate and
the electric field giving rise to ionization waves (striations).
In the next section, we will calculate the time of trapping with a
simple Brownian model.

\begin{table}\label{Table2}
  \centering
  \caption{Minimum electric field at reversal point and well width.
  Conditions: He gas, $p=1$ Torr, $l_0=20$ cm, $V_a=1$ kV, $\frac{T_{ew}}{T_{ep}}=0.1$, $\frac{n_{ew}}{n_{ep}}=10$.}
  \begin{tabular}{|c|c|}
    \hline 
    $E_m$ (V.cm$^{-1}$) & $\Delta \xi$ (cm) \\
    \hline
    $\lesssim -2.5$  & $2.6 \times 10^{-3}$ \\
    \hline
  \end{tabular}
\end{table}

From Eq.~\ref{Eq8} we calculate the cathode fall length for some
gases. For this purpose we took He and H$_2$ data as reference for
atomic and molecular gases, resp. The orders of magnitude are the
same, with the exception of Ar. Due to Ramsauer effect direct
comparison is difficult.

In Table III it is shown a comparison of the experimental cathode
fall distances to the theoretical prediction, as given by
Eq.~\ref{Eq16}. Taking into account the limitations of this model
these estimates are well consistent with experimental
data~\cite{Brown59}.

\begin{table}
  \centering
  \caption{Comparison between theoretical and experimental cathode fall distance at p=1 Torr,
  $E/p$=100 V$/$cm$/$Torr. Experimental data are collected from Ref.~\cite{Brown59}. }\label{Table2}
\begin{tabular}{|c|c|c|}
  \hline
  Gas & $\xi_c^{teo}$ (cm) & $\xi_c^{exp}$ (cm) \\
  \hline
  Ar    & 7.40      & 0.29 (Al)  \\
  He    & 1.32     & 1.32 (Al) \\
  $H_2$ & 0.80      & 0.80  (Cu) \\
  $N_2$ & 0.45     & 0.31 (Al) \\
  $Ne$  & 0.80      & 0.64 (Al) \\
  $O_2$ & 0.30      & 0.24 (Al) \\
  \hline
\end{tabular}
\end{table}

\subsection{Lifetime of a slow electron in the potential well}

The trapped electrons most probably diffuse inside the well with a
characteristic time much shorter than the lifetime of the through.
Trapping can be avoided by Coulomb collisions~\cite{Nishikawa69}
or by the ion-wave instability, both probably one outcome of the
stress energy unbalance as previously mentioned. We consider a
simple Brownian motion model for the slow electrons to obtain the
scattering time $\tau$, and the lifetime T of the well. A
Fermi-like model will allow us to obtain the slow electron energy
distribution function.

Considering the slow electron jiggling within the well, the
estimated scattering time is
\begin{equation}\label{scat1}
\tau = \frac{(\Delta \xi)^2}{ \mathcal{D}_e }.
\end{equation}
Here, $\mathcal{D}_e$ is the electron diffusion coefficient at
thermal velocities.

\begin{table*}
  \caption{Scattering time and trapping time in the well. The parameters are:
  $E/N=100$ Td, $T_g=300$ K, $V_a=1$ kV and $l_0=0.1$ m.}\label{Table 4}
  \begin{ruledtabular}
  \begin{tabular}{cccccc}
    Gas & $\mathcal{D}_e$ (cm$^2$.s$^{-1}$)\footnote{Data obtained through resolution of the homogeneous electron Boltzmann equation
    with two term expansion of the distribution function in spherical harmonics, M. J. Pinheiro and J. Loureiro,
    J. Phys. D.: Appl. Phys. {\bf 35} 1 (2002) }  & $\nu_{enw} (s^{-1})$ \footnote{Same remark as in ${}^a$} & $\Delta \xi (cm)$ & $\tau$ (s) & T (s) \\
    \hline
    Ar    & $2.52 \times 10^6$  & $8.10 \times 10^9$ & $1.34 \times 10^{-3}$  & $7.10 \times 10^{-13}$ & $3.97 \times 10^{-5}$ \\
    He    & $5.99 \times 10^6$  & $2.39 \times 10^9$ & $1.54 \times 10^{-2}$ & $3.95 \times 10^{-11}$ & $1.70 \times 10^{-5}$\\
    N$_2$ & $6.11 \times 10^5$  & $6.15 \times 10^9$ & $2.32 \times 10^{-3}$ & $8.81 \times 10^{-12}$  &$1.64 \times 10^{-4}$  \\
    CO$_2$ & $1.70 \times 10^6$ & $3.60 \times 10^9$ & $6.78 \times
    10^{-3}$& $2.70 \times 10^{-11}$ & $5.90 \times 10^{-5}$ \\
  \end{tabular}
  \end{ruledtabular}
\end{table*}

The fluctuations arising in the plasma are due to the breaking of
the well and we can estimate the amplitude of the fluctuating
field by means of Eq.~\ref{Eq14}. We obtain
\begin{equation}\label{}
\delta E_m =
\frac{n_{ep}}{n_{ew}}\frac{\nu_{enw}^2}{\nu_{en}^2}\frac{V_a}{e
l_0^2} \Delta \xi.
\end{equation}

Then, we have
\begin{equation}\label{}
\mathcal{E}_c = \frac{\delta E_m}{E_m} = \frac{\Delta \xi}{l_0}.
\end{equation}

In Table IV we summarize scattering and trapping times for a few
gases.

\subsection{Power-law slow electrons distribution function}

As slow electrons are trapped by the electric field inversion,
some process must be at work to pull them out from the well. We
suggest that fluctuations of the electric field in the plasma
(with order of magnitude of $\mathcal{E}_c$)act over electrons
giving energy to the slow ones, which collide with those
irregularities as with heavy particles. From this mechanism it
results a gain of energy as well a loss. This model was first
advanced by E. Fermi~\cite{Fermi} when developing a theory of the
origin of cosmic radiation. We shall focus here on the rate at
which energy is acquired.

The average energy gain per collision by the trapped electrons (in
order of magnitude) is given by
\begin{equation}\label{}
\Delta w = \overline{U} w(t),
\end{equation}
with $\overline{U}\cong \mathcal{E}_c^2$ and where $w$ is their
kinetic energy. After $N$ collisions the electrons energy will be
\begin{equation}\label{}
w(t) = \varepsilon_{t} \exp \left( \frac{\overline{U}t}{\tau}
\right),
\end{equation}
with $\varepsilon_t$ being their thermal energy, typical of slow
electrons.
\begin{figure}
  \includegraphics[width=3 in, height=4.5 in]{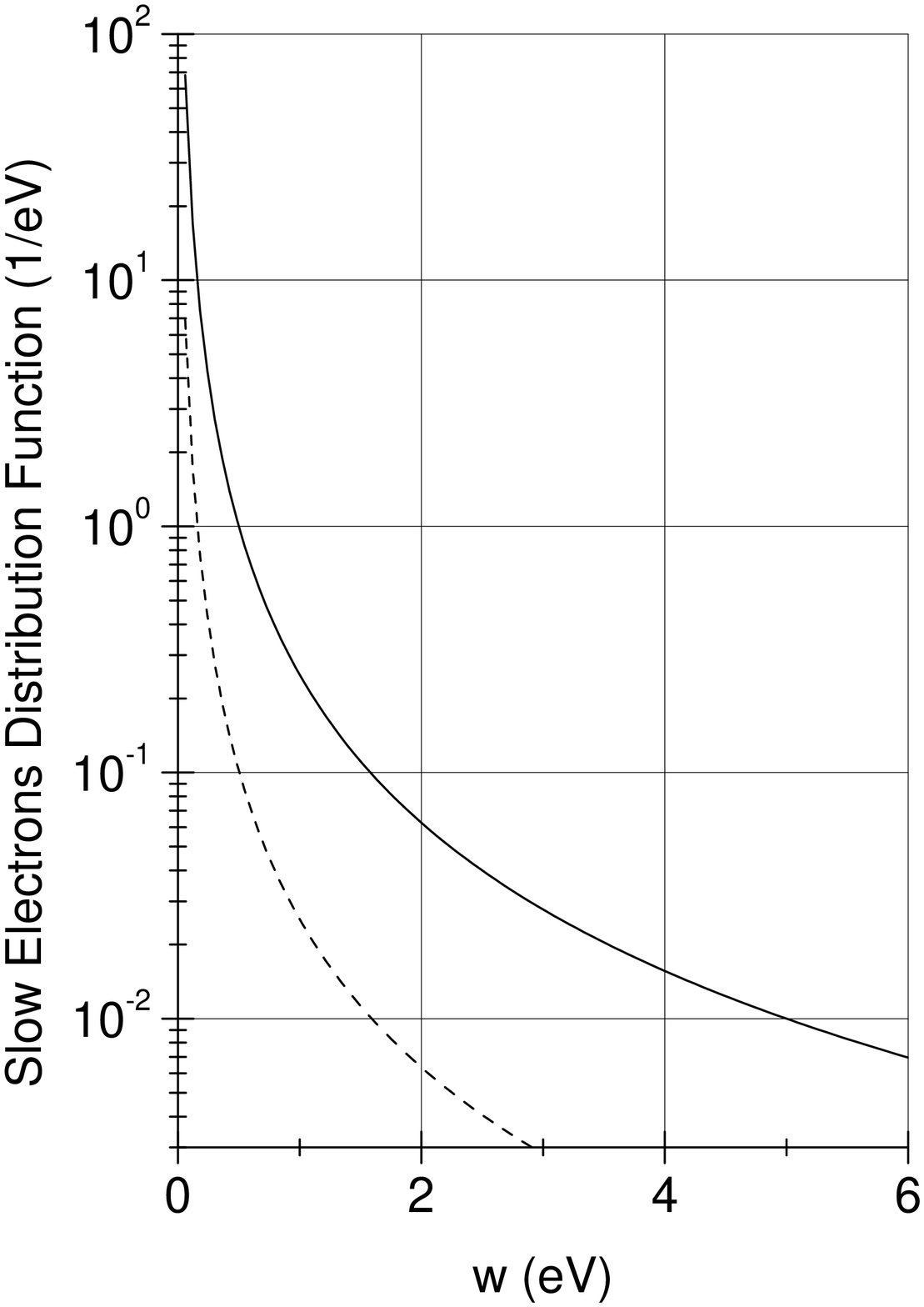}\\
  \caption{Slow electrons distribution function vs. energy, for the same conditions as presented in Table 4. Solid curve: Ar, broken curve: N$_2$.}\label{}
\end{figure}
The time between scattering collisions is $\tau$. Assuming a
Poisson distribution $P(t)$ for electrons escaping from the
trapping, then we state
\begin{equation}\label{5}
  P(t)=\exp(-t/\tau)dt/T.
\end{equation}
The probability distribution of the energy gained is a function of
one random variable (the energy), such as
\begin{equation}\label{6}
  f_w(w)d w = P\{ w<\bar{w}<w+dw \}.
\end{equation}
This density $f_w(w)$ can be determined in terms of the density
P(t). Denoting by $t_1=T$ the real root of the equation
$w=w(t_1=T)$, then it can be readily shown that slow electrons
obey in fact to the following power-law distribution function
\begin{equation}\label{7}
  f_w(w) d w = \frac{\tau}{\bar{U}T}
  \varepsilon_{t}^{\frac{\tau}{\bar{U}T}} \frac{d
  w}{w^{1+\tau/\bar{U}T}}.
\end{equation}
Like many man made and naturally occurring phenomena (e.g.,
earthquakes magnitude, distribution of income), it is expected the
trapped electron distribution function to be a power-law (see
Eq.~\ref{7}), hence $1 + \frac{\tau}{\mathcal{E}_c^2 T} = n$, with
$n=2 \div 4$ as a reasonable guess. Hence, we estimate the
trapping time to be
\begin{equation}\label{traptime}
T \approx \frac{\tau}{\mathcal{E}_c^2 n}.
\end{equation}

Fig.1 shows the slow electrons distribution function pumped out
from the well for two cases: Ar (solid curve), and N$_2$ (broken
curve). It was chosen a power exponent $n=2$. Those distributions
show that the higher confining time is associated with less slow
electrons present in the well. When the width of the well
increases (from solid to broken curve) the scattering time become
longer, and as well the confining time, due to a decrease of the
relative number of slow electrons per given energy. This mechanism
of pumping out of slow (trapped) electrons from the well can
possibly explains the generation of electrostatic plasma
instabilities.

Note that the trapping time is, in fact, proportional to the
length of the NG and inversely proportional to the electrons
diffusion coefficient at thermal energies:
\begin{equation}\label{}
T \approx \frac{l_0^2}{\mathcal{D}_e}.
\end{equation}
The survival frequency of trapped electrons is $\nu_t=1/T$. As the
electrons diffusion coefficient are typically higher in atomic
gases, it is natural to expect plasma instabilities and waves with
higher frequencies in atomic gases. This result is in agreement
with a kinetic analysis of instabilities in microwave
discharges~\cite{Tatarova1}. In addition, the length of the NG
will influence the magnitude of the frequencies registered by the
instabilities, since wavelengths have more or less space to
build-up. Table~\ref{Table 4} summarizes the previous results for
some atomic and molecular gases. The transport parameters used
therefor where calculated by solving the electron Boltzmann
equation, under the two-term approximation, in a steady-state
Townsend discharge~\cite{Pinheiro}.

\section{Conclusion}

This paper reviews, and extends when necessary, the physics of the
electric field reversals in a glow discharge. We listed and
studied how the structure of a glow discharge is related to the
electrons and ions kinetics, described related experimental
results, and have shown how the field reversals help to maintain
the discharge. To this end different methods to describe the
different groups of electrons generated in a glow discharge have
been presented: the nonlocal approach of Kolobov and
Tsendin~\cite{Kolobov:92}; the simple analytic fluid model by
Boeuf and Pitchford~\cite{Boeuf}; the theoretical analysis
developed by Kudryavtsev and Toinova~\cite{Kudryavtsev:05}, and
the dielectric-like model~\cite{Pinheiro:04}.


\bibliographystyle{amsplain}
\bibliography{Doc2}
\end{document}